\documentclass[a4paper]{jpconf}
\usepackage{graphicx}
\usepackage{amsmath}
\usepackage{subcaption}

\begin{document}
\title{On wind farm wake mixing strategies using dynamic individual pitch control} 

\author{Joeri Frederik, Bart Doekemeijer, Sebastiaan Mulders, Jan-Willem~van~Wingerden}

\address{Delft Center of Systems \& Control (DCSC), Department of Mechanical, Maritime and Materials Engineering (3mE), Delft University of Technology, The Netherlands.}

\ead{j.a.frederik@tudelft.nl}

\begin{abstract}
Dynamic wind farm control is a new strategy that aims to apply time-varying, often periodic, control signals on upstream wind turbines to increase the wake mixing behind the turbine. As a result, wake recovery is accelerated, leading to a higher power production of downstream turbines. As the amount of interest in dynamic control strategies for wind turbines in a wind farm is increasing, different approaches are being proposed. One such novel approach is called Dynamic Individual Pitch Control (DIPC). In DIPC, each blade pitch angle of a turbine is controlled independently to dynamically manipulate the direction of the thrust force vector exerted on the wind. Hence, the direction of the wake is varied, inducing wake mixing without significant thrust force magnitude variations on the rotor. In this paper, the effectiveness of different variations in the thrust direction are evaluated and compared using Large Eddy Simulation (LES) experiments. 

\end{abstract}

\section{Introduction}\label{sec:intro}
In the quest to increase the power production of wind farms, researchers and companies are looking for control strategies that increase the power production of turbines located in the wake of another turbine. Such turbines traditionally have a significantly lower power production than their upstream counterpart. Until recently, the vast majority of research into decreasing this power deficit has focused on steady-state operation. Given certain inflow conditions, the optimal settings of either the induction or the yaw angle of each turbine is assumed to be constant. For example, strategies such as wake redirection by means of yaw misalignment have proven capable to substantially reduce this power deficit \cite{Campagnolo2016_WFC,fleming2017,howland}. 

A consequence of this approach is that time-varying control inputs that influence the inherently dynamic nature of wind are disregarded. An example of such a time-varying control input is a periodic variation of the induction, either through the thrust coefficient \cite{Munters:2018} or by means of collective pitching \cite{frederik2019}. This approach, called Dynamic Induction Control (DIC), induces wake mixing and is verified to increase wind farm power capture in both Large Eddy Simulations (LES) \cite{Munters:2018} and scaled wind tunnel experiments \cite{frederik2019}. On the other hand, as shown in \cite{frederik2019}, the loads on the turbines are expected to increase with this strategy.

In \cite{frederik2020}, an alternative approach using Dynamic Individual Pitch Control (DIPC) is proposed. This control technology uses Individual Pitch Control (IPC) to dynamically manipulate the \textit{direction} of the thrust force of a turbine, hence increasing downstream wake mixing. IPC is a well-established control strategy in wind turbines, used mostly to reduce periodic loading \cite{bossanyi2003individual,bossanyi2005further}. Often the Multi-Blade Coordinate (MBC) transformation is used \cite{mbc,mulders2019analysis}, although different, more complex control algorithms have also been developed \cite{navalkar2014subspace,frederik2018cep}. In \cite{ipcwakesteering}, it has been shown that IPC can also be used to (statically) steer the wake of a turbine, although \cite{fleming2015ipc} shows that the potential increase in power capture of this approach is limited.

The strategy proposed in \cite{frederik2020} combines wake redirection through IPC with dynamic control concepts. The individual blade pitch angles of an upstream turbine are manipulated in such a way that the direction of the wake is changed over time. This dynamic wake behavior leads to increased interaction with the free-stream wind flow, inducing enhanced wake recovery. It is shown in \cite{frederik2020} that this approach can lead to an energy increase in the wake of up to $47\%$, while the power of the controlled turbine decreases by less than $3\%$. Unlike strategies such as wake redirection through yaw and DIC, DIPC does not deviate from the operating range for which the turbine was designed. This approach may therefore be a more realistically implementable strategy to reliably increase the power production in existing wind farms.

In this paper, different DIPC strategies, dynamically redirecting the wake either in vertical or horizontal direction, or both, are evaluated and compared. This paper is organized as follows: Section~\ref{sec:control} describes the control strategy behind DIPC, followed by the definition of the simulation environment in Section~\ref{sec:sowfa}. The results of these simulations will be shown and evaluated in Section~\ref{sec:results}, before conclusions will be drawn in Section~\ref{sec:concl}.

\section{Control strategy}\label{sec:control}

In this section, the concept of DIPC, and how it can be implemented on a wind turbine, is explained. DIPC makes use of the notion that each blade has an individual contribution to the overall thrust force of a turbine on the wind. By varying the pitch angle of each blade depending on its azimuth angle, the thrust force of the turbine can be different at different locations of the rotor swept disk. 

To effectively manipulate this notion, the Multi-Blade Coordinate (MBC) transformations are used to project the individual blade moments onto a non-rotating frame. For three-bladed turbines, this transformation is given as

\begin{align}
\left[ \begin{array}{c} M_{0}(t) \\ M_{{\textrm{tilt}}}(t) \\ M_{{\textrm{yaw}}}(t) \end{array} \right] &= \mathbf{T}(\psi) \underbrace{\left[ \begin{array}{c} M_{1}(t) \\ M_{2}(t) \\ M_{3}(t) \end{array} \right]}_{M(t)}\label{eq:mbc},
\end{align}
with
\begin{align}
\nonumber\mathbf{T}(\psi) &= \frac{2}{3}\left[ \begin{array}{c c c} 0.5 & 0.5 & 0.5 \\ \cos{\left(\psi_1\right)} & \cos{\left(\psi_2\right)} & \cos{\left(\psi_3\right)} \\ \sin{\left(\psi_1\right)} & \sin{\left(\psi_2\right)} & \sin{\left(\psi_3\right)} \\ \end{array} \right],
\end{align}
where $\psi_b$ is the azimuth angle of blade $b$, with $\psi_b = 0$ indicating the vertical upright position. The collective moment $M_0$ represents the cumulative rotor moment perpendicular to the rotor swept disk, while $M_{\text{tilt}}$ and $M_{\text{yaw}}$ are the fixed-frame vertical and horizontal moment, respectively. By applying the inverse MBC transformation $T^{-1}(\psi)$, the implementable individual blade pitch angles can be obtained from the fixed-frame pitch angles in a similar fashion. 

\begin{figure}[t]
    \centering
    \includegraphics[width=\textwidth]{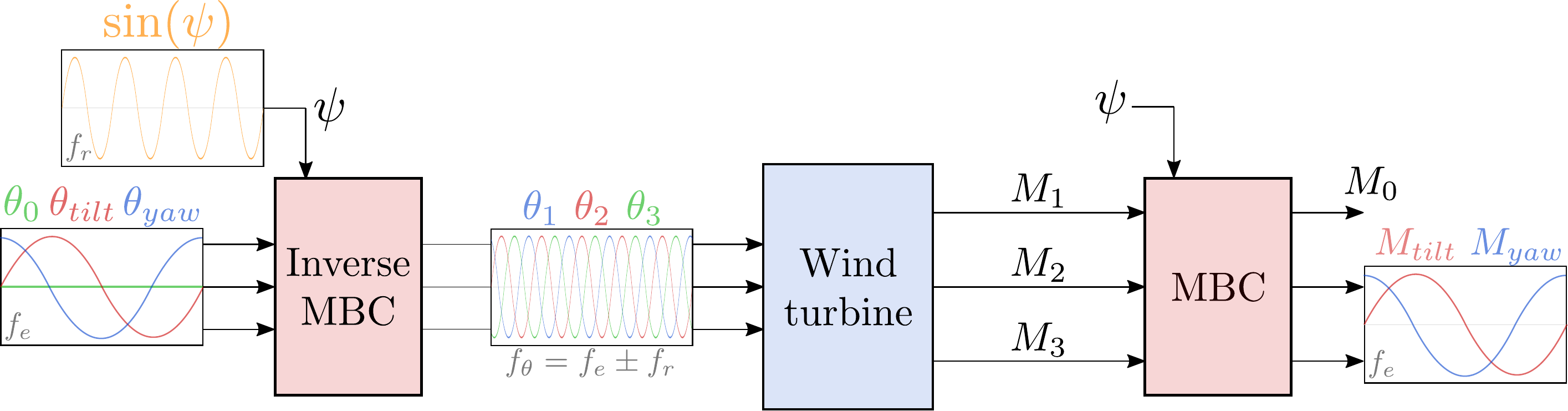}
    \caption{The control scheme used for DIPC of a wind turbine, as shown in \cite{frederik2020}. The individual pitch angles $[\theta_1 \,\, \theta_2 \,\, \theta_3]$ are obtained by applying the inverse MBC transformation on the desired tilt and yaw angles $\theta_{\mathrm{tilt}}$ and $\theta_{\mathrm{yaw}}$. In this example, a periodic excitation on both the tilt and yaw angle leads to a similar periodic variation of the tilt and yaw moments $M_{\mathrm{tilt}}$ and $M_{\mathrm{yaw}}$.}
    \label{fig:blockscheme}
\end{figure}

The DIPC strategy revolves around creating a dynamically varying tilt and/or yaw moment, in order to manipulate the vertical and/or horizontal direction of the wake, respectively, over time. By defining the tilt and/or the yaw angle as a low-frequent sinusoid, tilt and/or yaw moments $M_{\mathrm{tilt}}$ and $M_{\mathrm{yaw}}$ can be achieved. The control scheme of this approach is shown in Figure~\ref{fig:blockscheme}.

Notice that the low-frequent sinusoid on the tilt and yaw angle leads to a higher-frequent sinusoid on the actual individual pitch angles. The frequency of the pitch excitation actually differs only slightly from the rotation frequency $1P$ that is commonly used for load alleviation IPC. As shown in \cite{frederik2020}, the pitch frequency $f_{\theta}$ equals
\begin{equation}
    f_{\theta} = f_r \pm f_e,
    \label{eq:pitchfreq}
\end{equation}
where $f_r$ is the rotation frequency and $f_e$ the excitation frequency applied on the tilt or yaw angle. Note that the optimal excitation frequency is not exactly known yet (see \cite{frederik2020}). Therefore, the assumption is made that this optimum is similar to the optimum in DIC, which is around a Strouhal number of $St = 0.25$ \cite{frederik2019,Munters:2017}. This is indeed results in a frequency $f_e$ that is significantly lower than $f_r$. For the NREL 5~MW reference turbine used in the simulations that will be introduced in Section~\ref{sec:sowfa}, the rated rotor speed is 12.1 rpm, i.e., $f_r \approx 0.20$ Hz. For the given Strouhal number,
\begin{equation}
    f_e = \frac{St U}{D} \approx 1.6\cdot10^{-2},
    \label{eq:exfreq}
\end{equation}
where $U = 8$ m/s is the free-stream wind velocity and $D = 126$ m is the rotor diameter. Clearly, the excitation frequency is an order of magnitude 10 smaller than the rotation frequency.

In Figure~\ref{fig:blockscheme}, a periodic excitation on both the tilt and the yaw angle is applied, leading to the \textit{helix approach} as described in \cite{frederik2020}, either in counterclockwise (CCW) or clockwise (CW) direction. It is found in \cite{frederik2020} that the CCW helix is substantially more effective than the CW helix. Therefore, only the CCW helix is considered in this paper.

\begin{figure}[t!]
    \centering
    \begin{subfigure}{0.23\textwidth}
    \includegraphics[width=\textwidth]{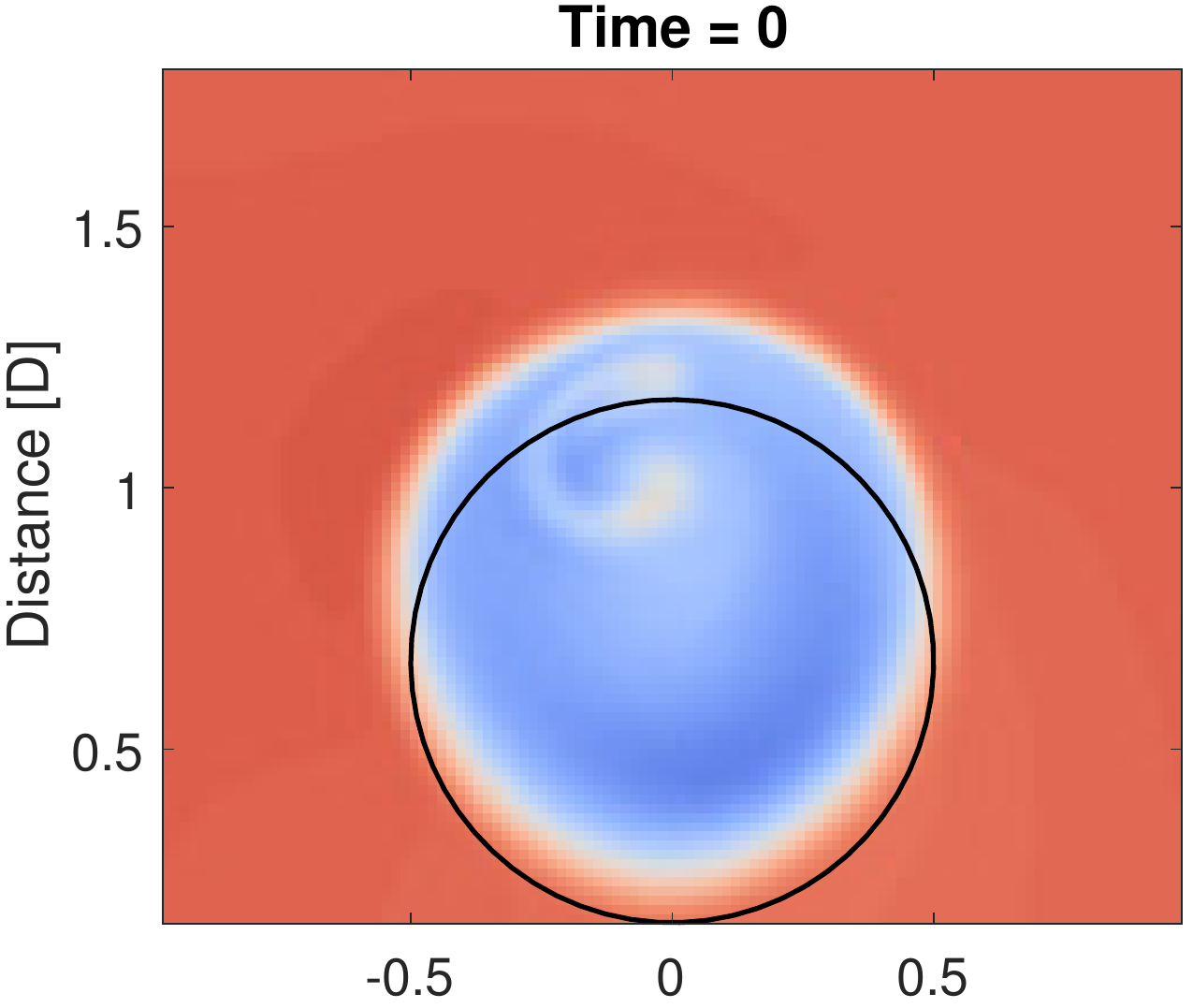}
    \end{subfigure}
    ~
    \begin{subfigure}{0.23\textwidth}
    \includegraphics[width=\textwidth]{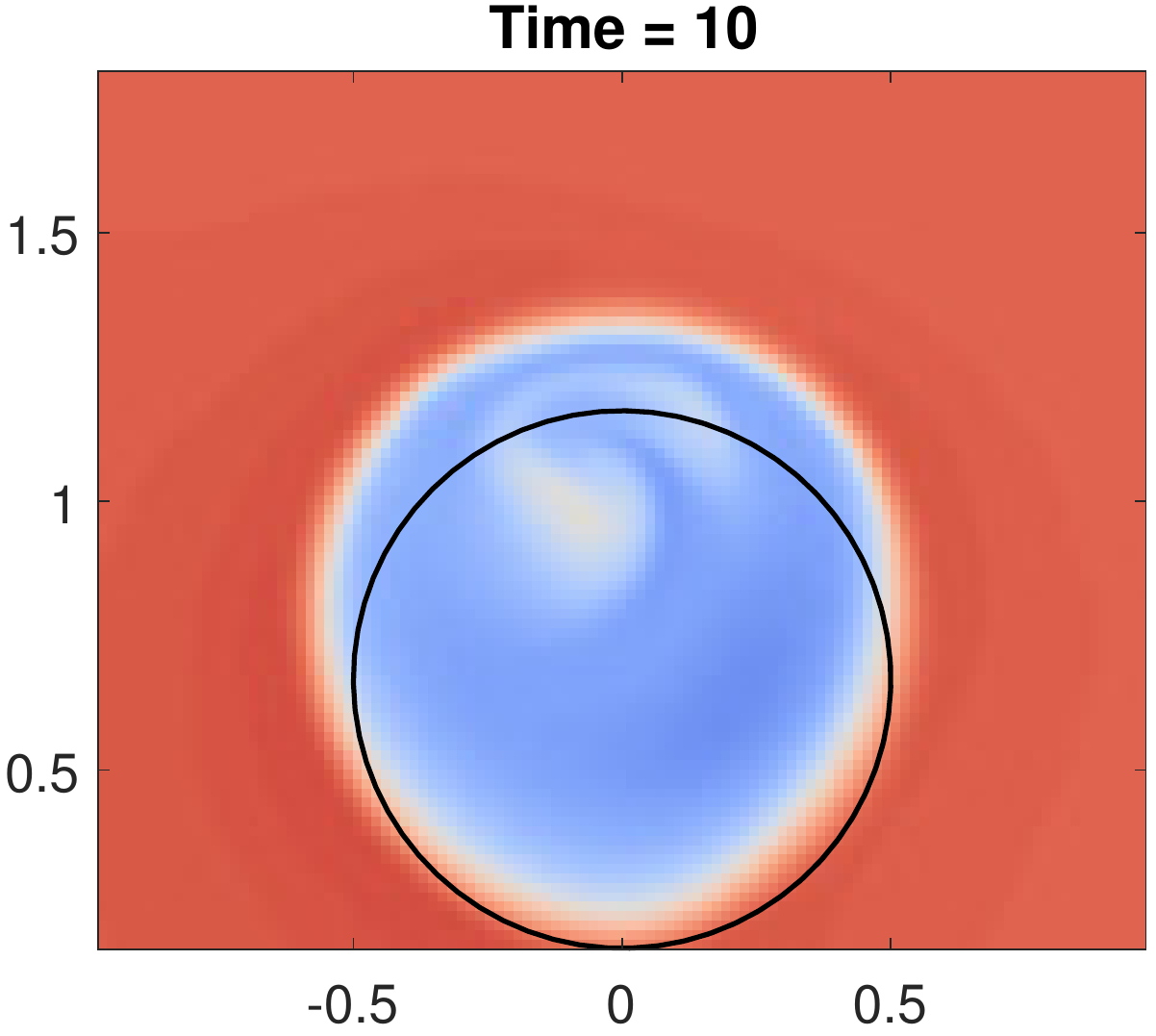}
    \end{subfigure}
    ~
    \begin{subfigure}{0.23\textwidth}
    \includegraphics[width=\textwidth]{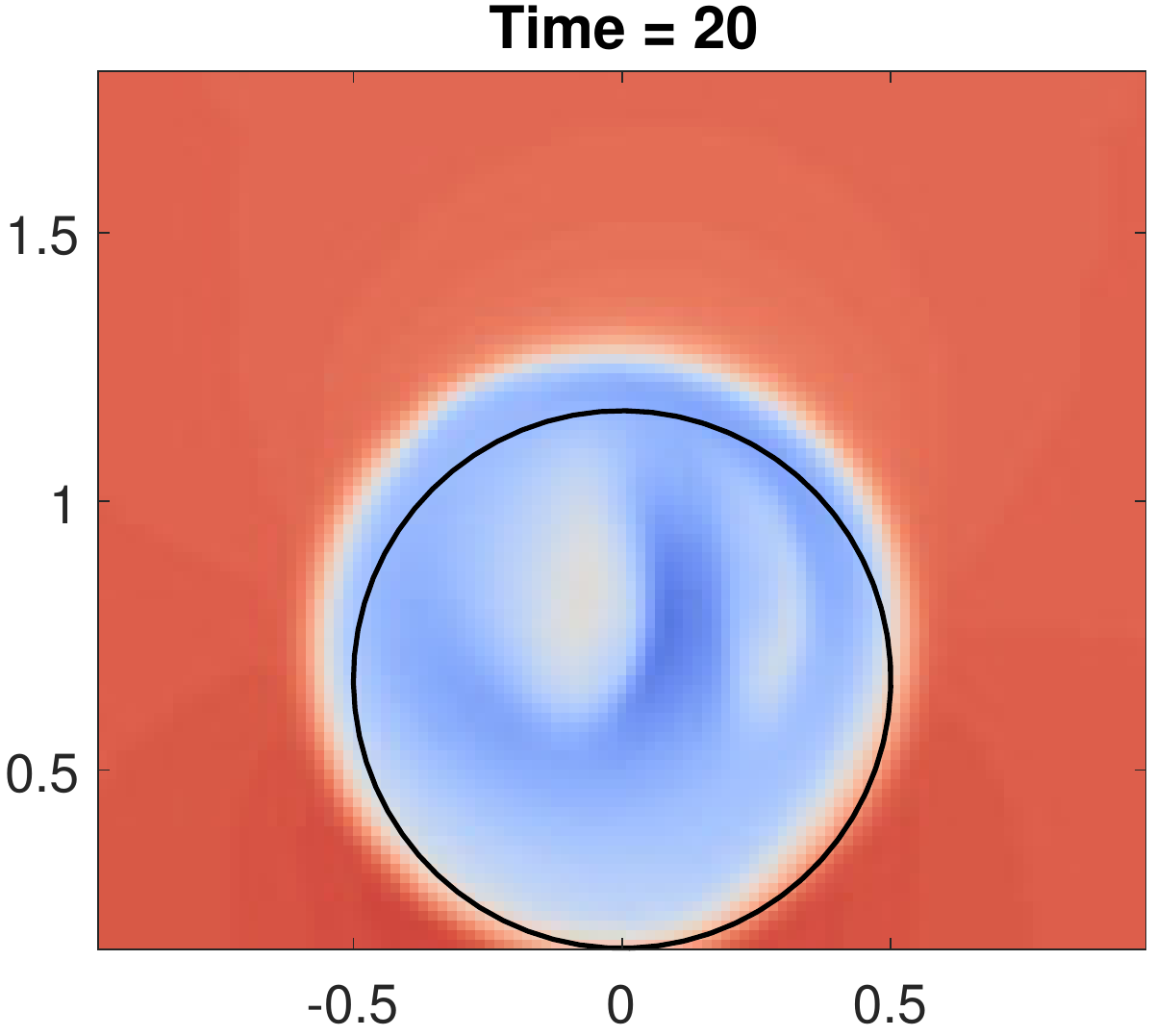}
    \end{subfigure}
    ~
    \begin{subfigure}{0.23\textwidth}
    \includegraphics[width=\textwidth]{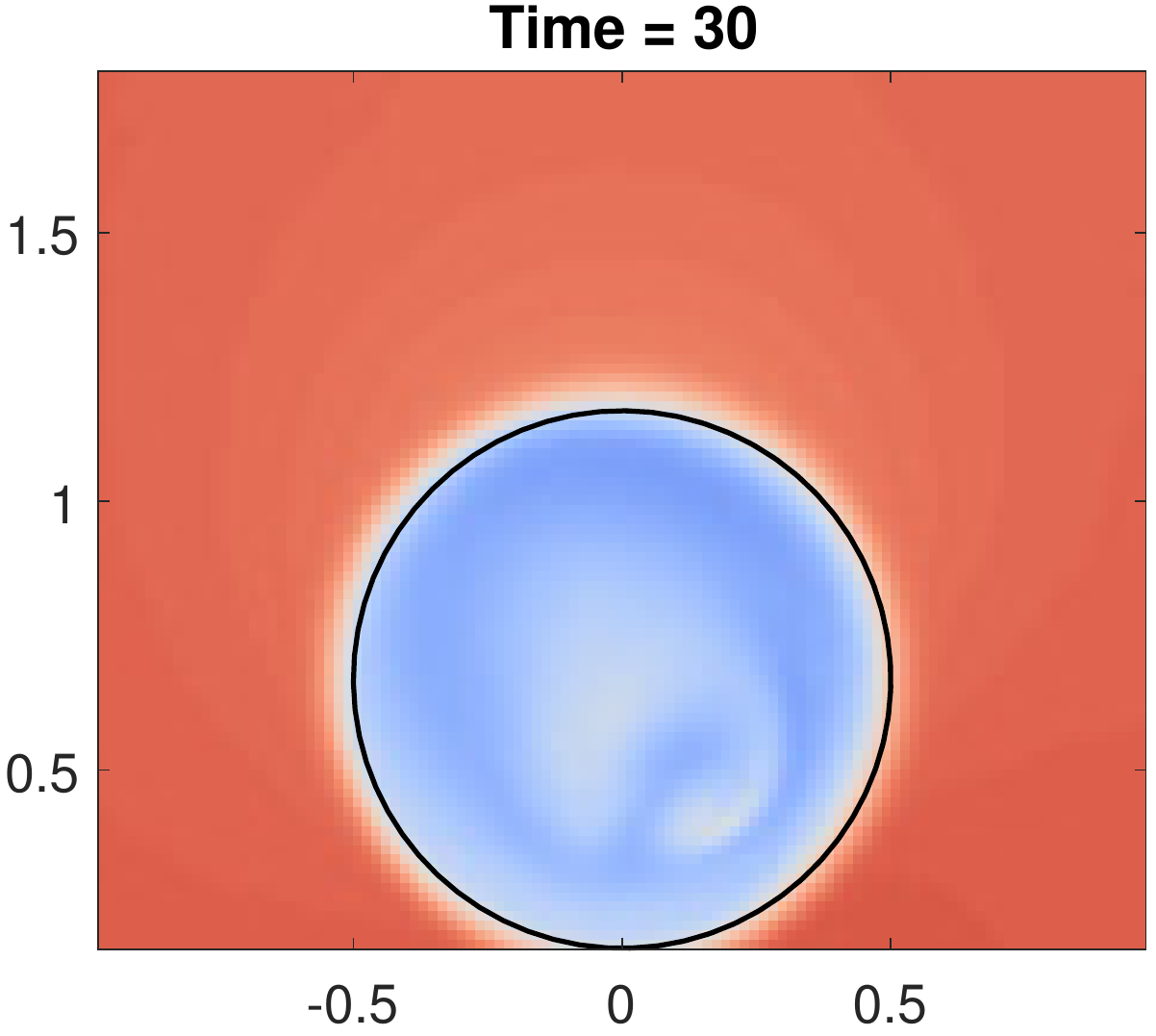}
    \end{subfigure}
    
    \begin{subfigure}{0.23\textwidth}
    \includegraphics[width=\textwidth]{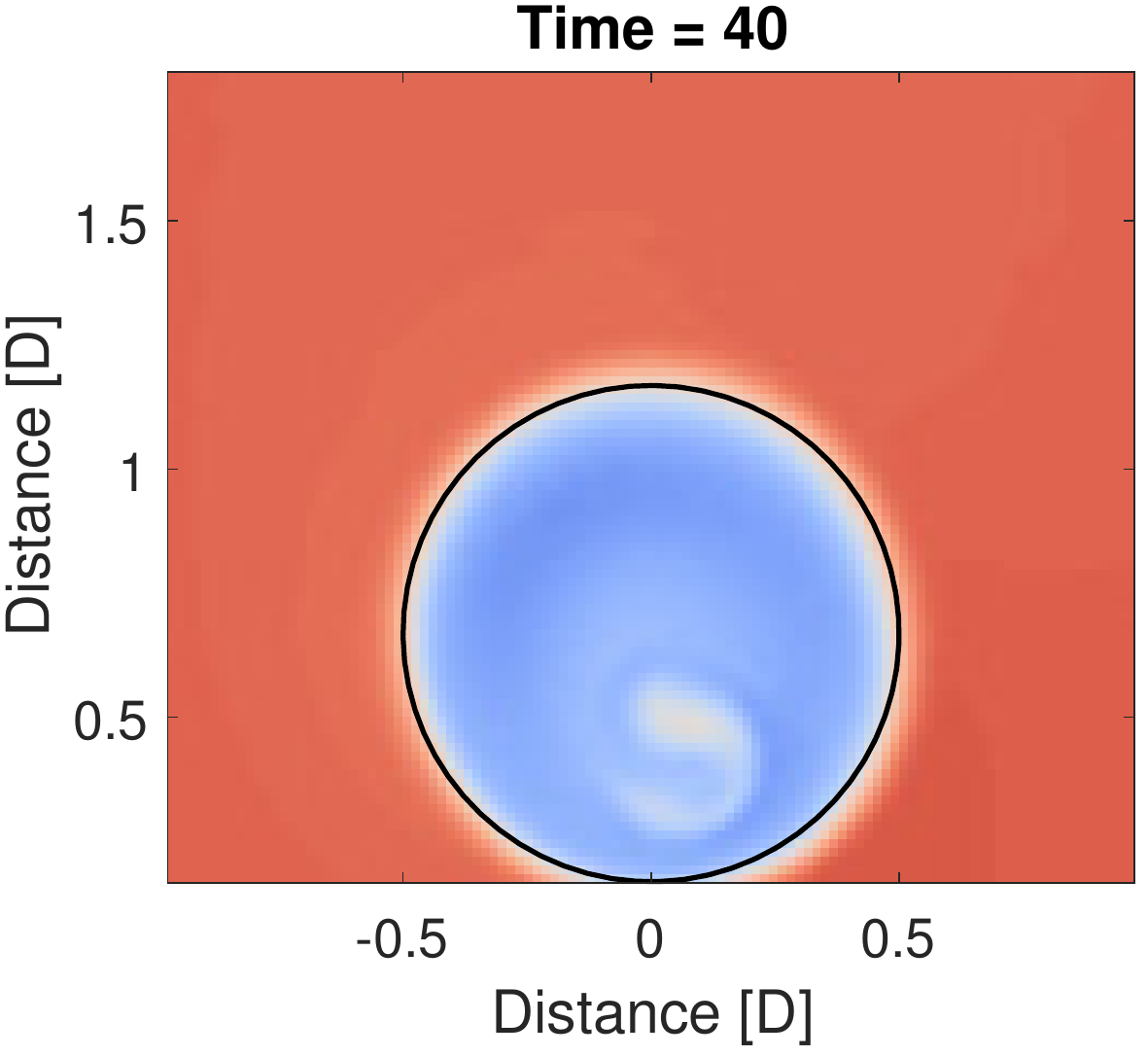}
    \end{subfigure}
    ~
    \begin{subfigure}{0.23\textwidth}
    \includegraphics[width=\textwidth]{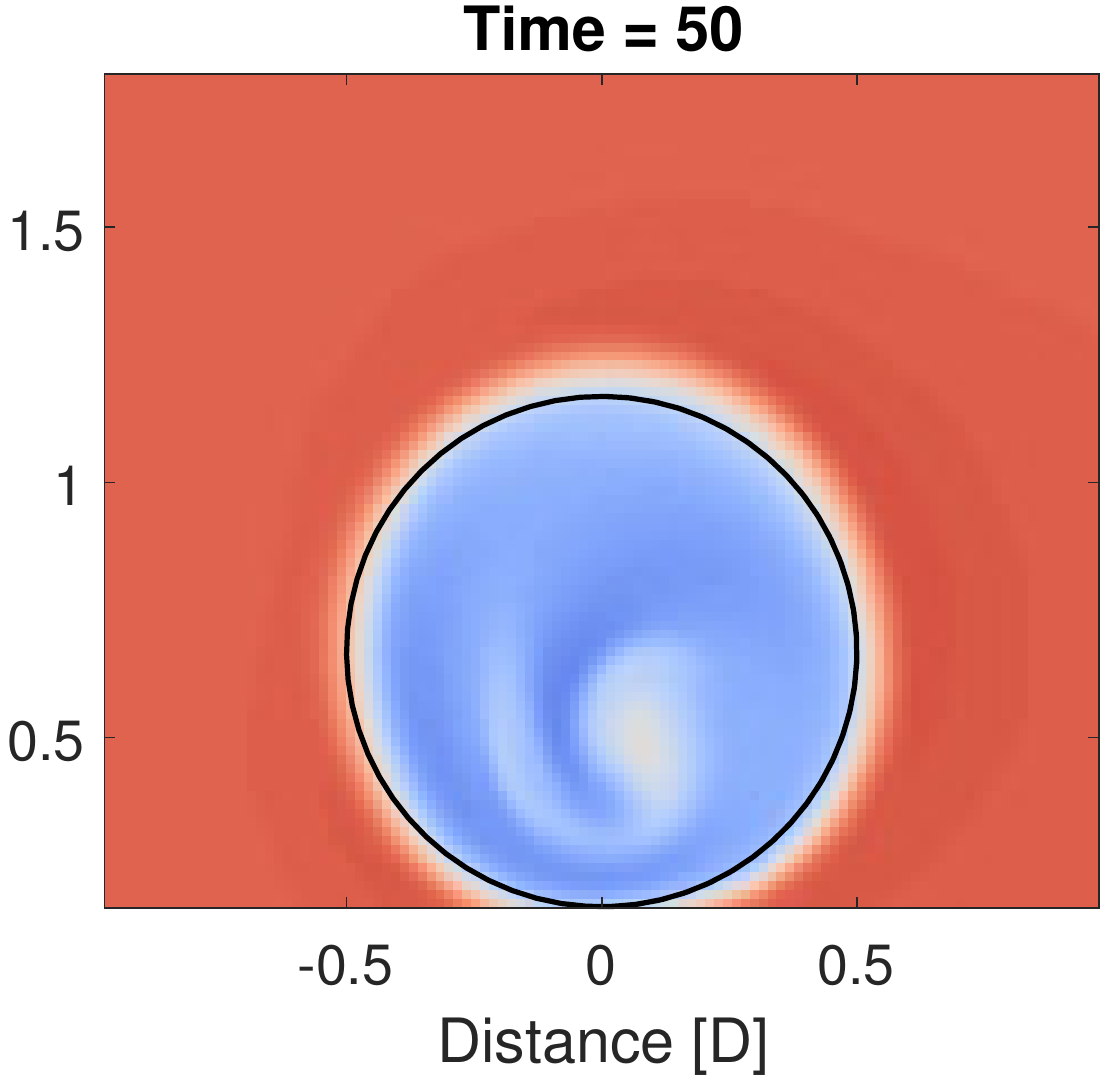}
    \end{subfigure}
    ~
    \begin{subfigure}{0.23\textwidth}
    \includegraphics[width=\textwidth]{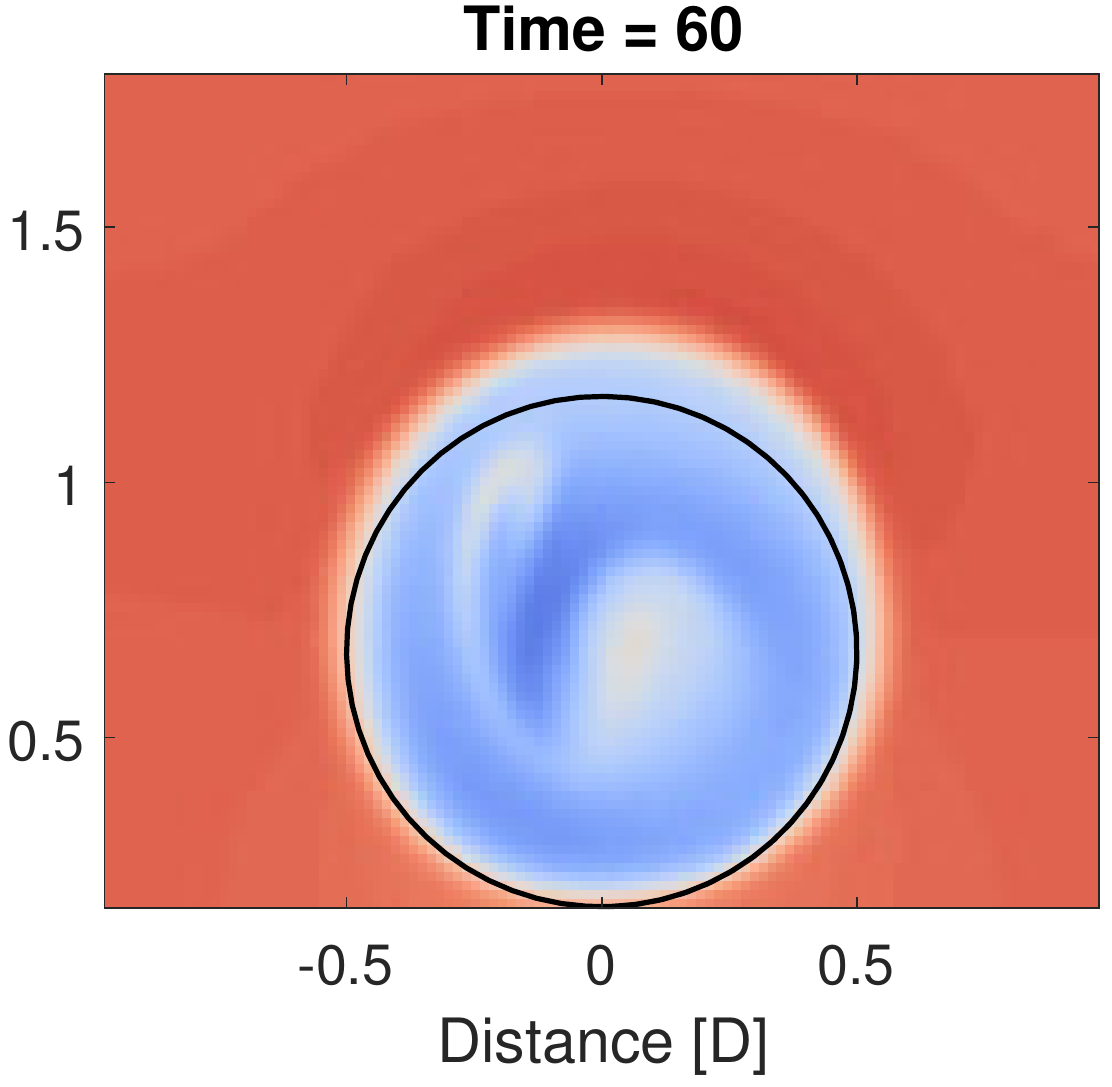}
    \end{subfigure}
    ~
    \begin{subfigure}{0.23\textwidth}
    \includegraphics[width=\textwidth]{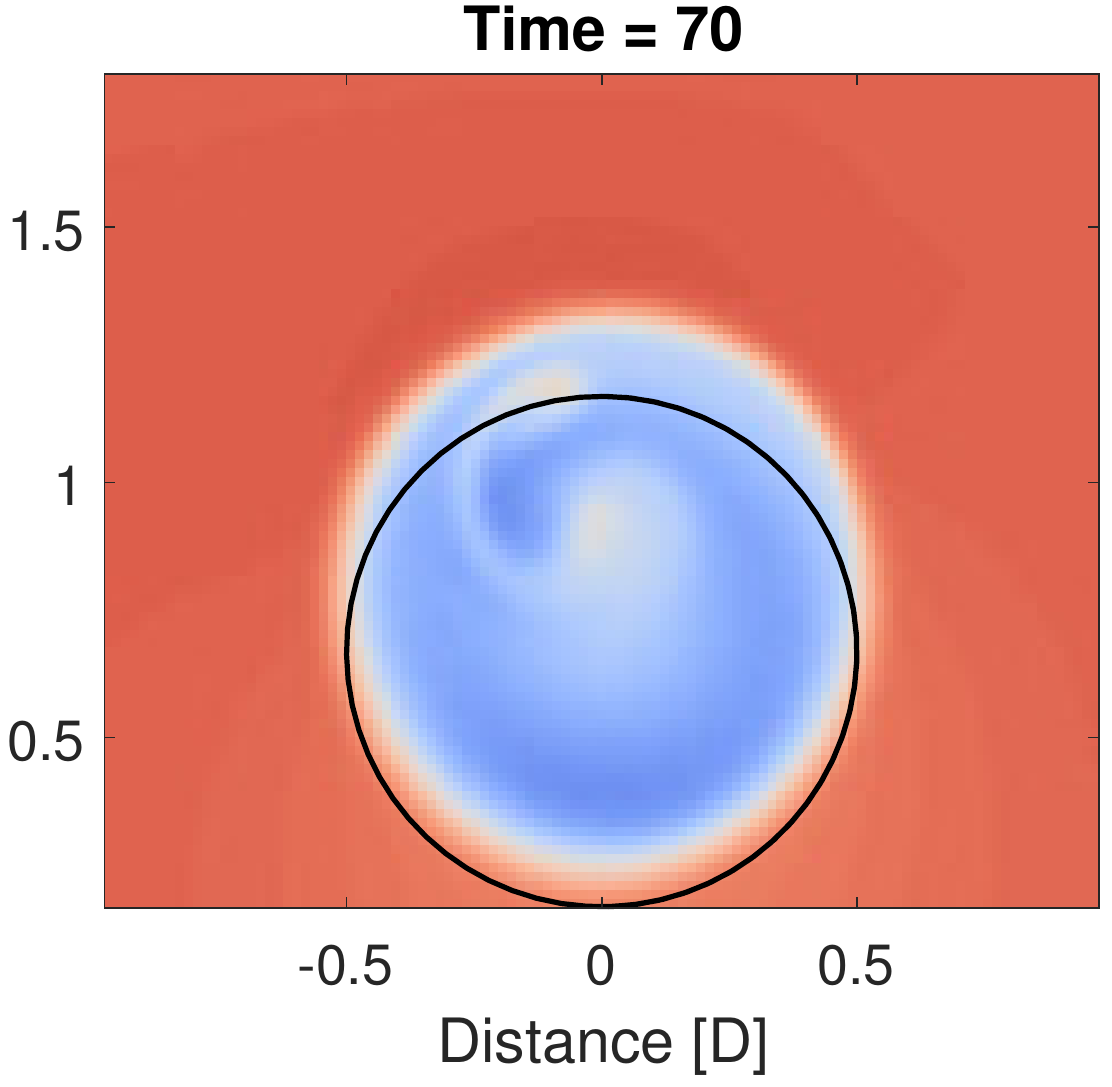}
    \end{subfigure}
    \caption{A wake as measured at 3$D$ behind the turbine at different time instances when a low-frequent sinusoid is imposed on the tilt angle of the upstream turbine (tilt DIPC). Red indicates a region of high wind velocity, blue indicates low wind velocity. Clearly, the wake is moving in vertical direction over time. Obtained using uniform inflow simulations in SOWFA.}
    \label{fig:waketilt}
\end{figure}

It is also possible to impose an excitation on one of these two angles, such that the wake direction is only varied either vertically (tilt DIPC) or horizontally (yaw DIPC). The former is visualized using simulations with a laminar wind inflow profile in Figure~\ref{fig:waketilt}. The effectiveness of such an approach is assessed in this paper, and compared to the helix approach. This results in three different DIPC strategies: yaw DIPC, tilt DIPC and the CCW approach. In all cases, the amplitude of the pitch excitation will be 2.5 degrees. The approaches are compared with \textit{baseline} greedy control -- where the turbine operates at static maximal power capture -- and static derating. 

The assesment of the different control strategies will be executed using high-fidelity simulations in the LES-code Simulator fOr Wind Farm Applications (SOWFA) \cite{SOWFA_General}, using a single 5~MW NREL reference turbine. The details of these simulations will be described in Section~\ref{sec:sowfa}.

The power capture of the controlled turbine will be investigated, as well as the energy available in the wake at different distances behind the turbine. This gives a solid indication of the power capture of potential downstream turbines at particular locations. Finally, the variation of the thrust force will be investigated as indicator of the loads on the turbine with different control strategies.

\section{Simulation environment}\label{sec:sowfa}

The control strategies described in Section~\ref{sec:control} are evaluated using high-fidelity simulations executed in SOWFA \cite{SOWFA_General}. SOWFA is a high-fidelity simulation environment developed by the National Renewable Energy Laboratory (NREL). The simulations are performed with a single NREL~5MW reference turbine \cite{NREL5MW}, using an inflow velocity of 8~m/s and a turbulence intensity of 5.9\%. The details of the simulation settings are shown in Table~\ref{tab:SOWFA}.

\begin{table}[h]
    \centering
    \caption{Numerical simulation scheme in SOWFA for uniform simulations}
    \label{tab:SOWFA}    
    \begin{tabular}{r l}
         & \\
         \textbf{Turbine} & NREL 5MW \cite{NREL5MW} \\
         \textbf{Rotor diameter} & $126.4$~m\\
         \textbf{Domain size} & $3$~km~$\times~3$~km~$\times$~$1$~km \\
         \textbf{Cell size (outer region)} & $10$~m~$\times~10$~m~$\times$~$10$~m \\
         \textbf{Cell size (near rotor)} & $1.25$~m~$\times~1.25$~m~$\times$~$1.25$~m\\
         \textbf{Inflow wind speed} & $8.0$~m/s  \\      
         \textbf{Inflow turbulence intensity} & $5.9\%$\\
    \end{tabular}
\end{table}

All simulations are executed with one turbine, which applies the control strategies described in Section~\ref{sec:control}. The wake behind the turbine is studied to determine the effect of the different approaches on the flow behind the turbine. Slices at integer rotor diameters $D$ are investigated to see how effective DIPC is at decreasing the wake velocity deficit. All results will be normalized with respect to the baseline case of greedy control, where the turbine operates at its individual steady-state optimum. These results are presented in Section~\ref{sec:results}.

\section{Results}\label{sec:results}

In this section, the results of the simulations in SOWFA as described in the previous section are discussed. For straightforward comparison, the results are normalized with respect to the baseline case of greedy control. 

\begin{figure}[t!]
    \centering
    \includegraphics[width=0.8\textwidth]{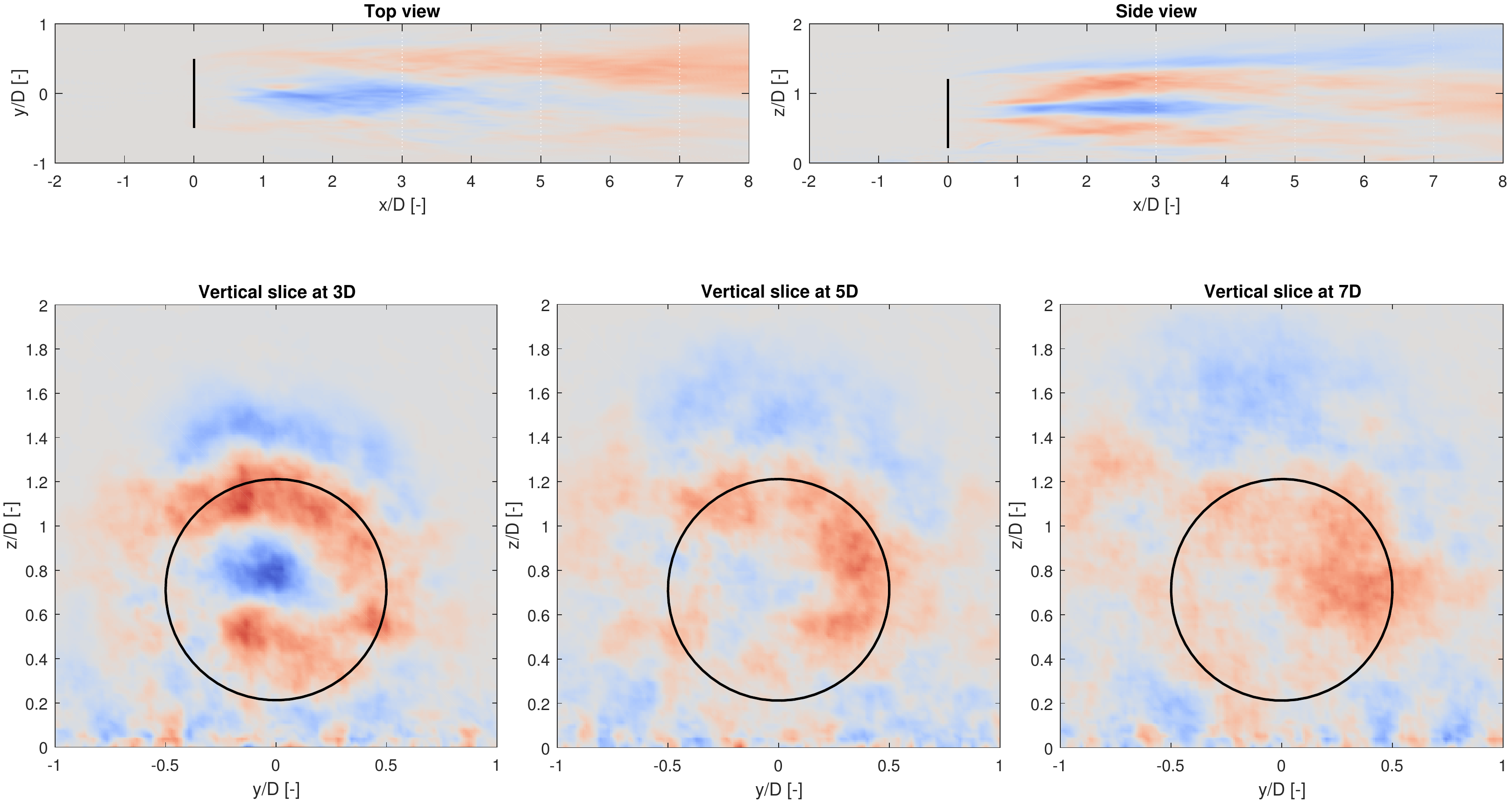}
    \caption{The mean wind speed, relative to the baseline case, at different distances behind the excited turbine, when tilt DIPC is applied. The red area's indicate that the flow velocity in the wake is increased, blue area's indicate a velocity decrease.}
    \label{fig:flowtilt}
\end{figure}

Figure~\ref{fig:flowtilt} shows the mean wind velocity around the turbine, when tilt DIPC is applied. This figure indicates that the velocity at the side of the blade swept disk area is substantially increased, while in the middle it is decreased. Further downstream, wake mixing reduces this difference in wind velocity. The average energy in the wake streamtube is increased with approximately $5-6\%$.

\begin{table}[b!]
    \centering
    \caption{Summary of the simulation concerning the wake behind the turbine. The energy results are given with respect to the baseline greedy control case.}
    \begin{tabular}{l|c|c|c|c|c}
         & \rotatebox[origin=c]{90}{\textbf{Yaw DIPC}} & \rotatebox[origin=c]{90}{\textbf{Tilt DIPC}} & \rotatebox[origin=c]{90}{\textbf{\,CCW Helix}} &
         \rotatebox[origin=c]{90}{\textbf{DIC}} &
         \rotatebox[origin=c]{90}{\textbf{SIC}}  \\
      \textbf{Energy at 3$D$} & +3.7\% & +5.2\% & +10.7\% & +18.6\% & +13.6\% \\\hline
        \textbf{Energy at 5$D$} & +2.7\% & +5.2\% & +10.7\% & +15.6\% & +7.1\% \\\hline
        \textbf{Energy at 7$D$} & +2.7\% & +6.2\% & +11.3\% & +12.4\% & +5.2\% \\
    \end{tabular}
    \label{tab:results}
\end{table}

\begin{figure}[t!]
    \centering
    \includegraphics[width=0.8\textwidth]{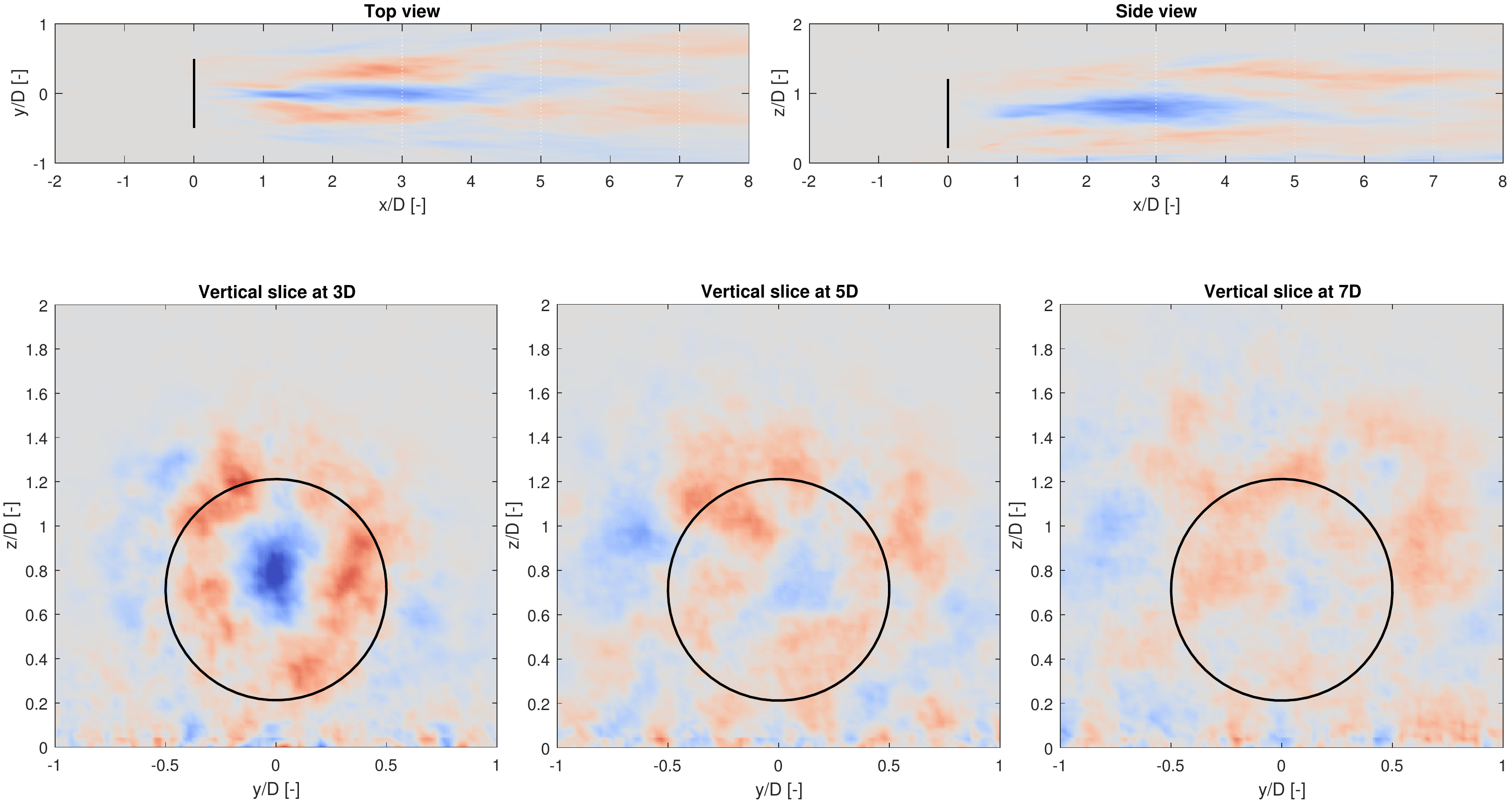}
    \caption{The mean wind speed, relative to the baseline case, at different distances behind the excited turbine, when yaw DIPC is applied. The red area's indicate that the flow velocity in the wake is increased, blue area's indicate a velocity decrease.}
    \label{fig:flowyaw}
\end{figure}

\begin{figure}[h!]
    \centering
    \includegraphics[width=0.8\textwidth]{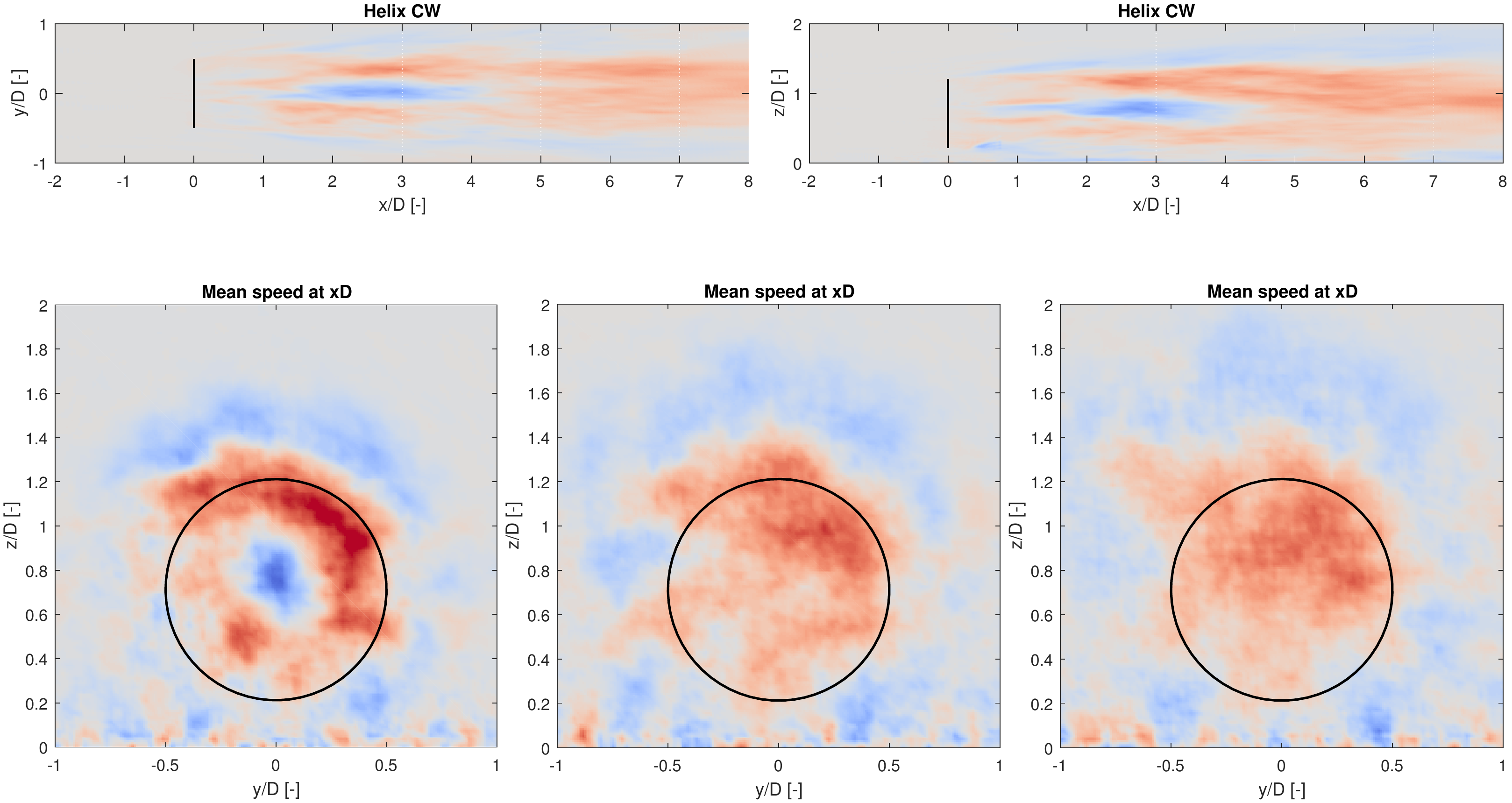}
    \caption{The mean wind speed, relative to the baseline case, at different distances behind the excited turbine, when the CCW helix method is applied. The red area's indicate that the flow velocity in the wake is increased, blue area's indicate a velocity decrease.}
    \label{fig:flowhelix}
\end{figure}

A similar analysis can be executed with yaw DIPC and the helix. The wake velocity of these strategies is shown in Figures~\ref{fig:flowyaw}~and~\ref{fig:flowhelix}, respectively. A comparison learns that the helix approach is clearly the most effective in enhancing wake mixing. The dissimilarities between the yaw and tilt DIPC are more subtle: based on the figures, it is hard to tell which method performs better. Data analysis shows that yaw DIPC leads to a wake energy increase of $2.5-4\%$, therefore performing slightly worse than tilt DIPC. These results are summarized in Table~\ref{tab:results}. For comparison, the results of DIC and static derating (Static Induction Control, SIC) are included as well.

\begin{figure}
    \centering
    \includegraphics[width=0.9\textwidth]{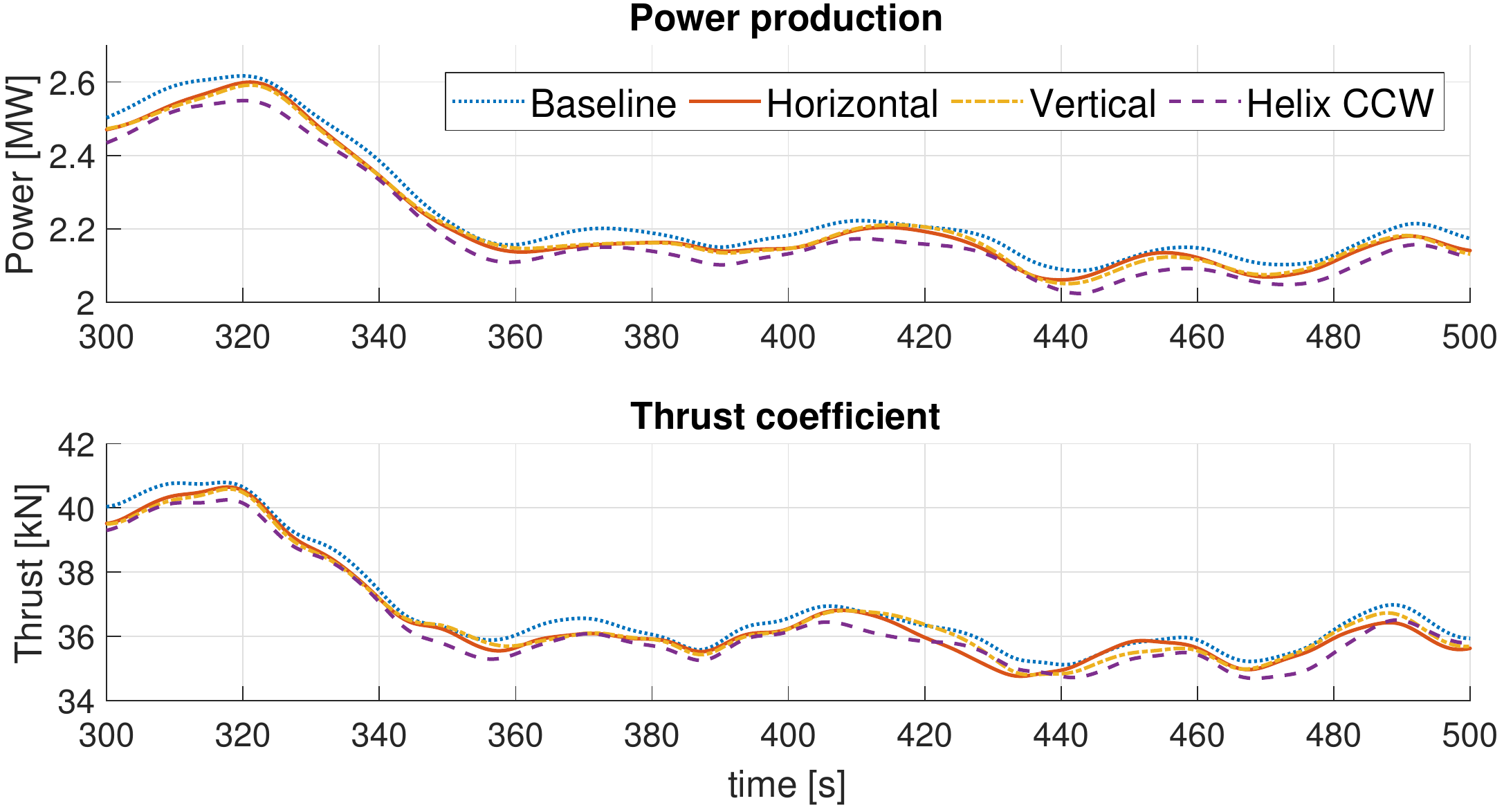}
    \caption{The power (top) and thrust (bottom) signals of the turbine with different control strategies. For cases, the power and thrust is slightly lower than baseline, with the helix method resulting in the most significant power loss.}
    \label{fig:powerthrust}
\end{figure}

Apart from the energy increase in the wake, it is of course interesting to investigate the effects of the different strategies on the performance of the turbine itself. This will be done through assessing the power capture of this turbine, as well as analyzing the thrust on the rotor disk. These signals are shown in Figure~\ref{fig:powerthrust} for the three different DIPC strategies as well as for the baseline case. This figure indicates that the power production of the tilt and yaw DIPC strategy is slightly higher than with the helix, albeit still lower than the baseline case. Clearly, this is the price to pay for the energy increase in the wake of the turbine.

The effect of the different control strategies on the turbine are summarized in Table~\ref{tab:results2}. This figure shows the mean and variance of the power, thrust and blade moment signals, as well as the pitch activity. These results show that yaw and tilt DIPC can be considered a "light" version of the helix approach: the wake mixing benefits are slightly lower, but so is the power loss, pitch activity and variations of blade moments. In all cases, the negative effects on the controlled turbine for DIPC are much lower than for DIC, while the power loss of static derating is substantially higher than for DIPC. This indicates that especially tilt DIPC, which performs slightly better than yaw DIPC, could be an interesting compromise for decreasing the wake deficit and with limited effects on the controlled turbine.

\begin{table}[h!]
    \centering
    \caption{Summary of the simulation results concerning turbine performance. All results, apart from the pitch activity, are given with respect to the baseline greedy control case.}
    \begin{tabular}{l|c|c|c|c|c}
         & \rotatebox[origin=c]{90}{\textbf{Yaw DIPC}} & \rotatebox[origin=c]{90}{\textbf{Tilt DIPC}} & \rotatebox[origin=c]{90}{\textbf{\,CCW Helix}} &
         \rotatebox[origin=c]{90}{\textbf{DIC}} &
         \rotatebox[origin=c]{90}{\textbf{SIC}}  \\
      \textbf{Power capture} & -1.2\% & -1.1\% & -2.3\% & -1.8\% & -4.0\%  \\\hline
        \textbf{Variance of power} & -3.0\% & -4.3\% & -6.5\% & +139.4\% & -9.4\% \\\hline
        \textbf{Thrust force} & -0.7\% & -0.6\% & -1.2\%  & -0.3\%& -8.6\% \\ \hline
        \textbf{Variance of thrust}& -3.3\% & -3.6\% & -5.0\% & +943.5\% & -18.0\% \\\hline
        \textbf{Moment on blades} & -0.8\% & -0.7\% & -1.4\% & +0.2\% & -9.2\% \\ \hline
        \textbf{Variance of moment} & +219.9\% & +224.4\% & +423.0\% & +659.1\% & -13.4\% \\ \hline
        \textbf{Pitch activity [deg/s]} & 0.99 & 0.99 & 1.69 & 0.16 & 0
    \end{tabular}
    \label{tab:results2}
\end{table}

\section{Conclusions}\label{sec:concl}
In this paper, different DIPC strategies are evaluated using the SOWFA environment. All strategies show to increase the wake recovery, demonstrating the potential of this approach as a viable method to increase the power production of wind farms. The results presented here indicate that wake manipulation in vertical direction, tilt DIPC, is more effective than its horizontal counterpart, yaw DIPC. Both approaches show better performance at the excited turbine than the helix approach, where the wake is manipulated both horizontally and vertically, but also prove less effective in decreasing the wake deficit.

The results presented here indicate that DIPC will lead to lower loads on the excited turbine than DIC. Compared to static induction, the power loss of the upstream turbine is much lower with DIPC, indicating that DIPC is a more effective method for wind farm power maximization. DIPC can therefore be considered a reasonable alternative to conventional, steady-state wind farm control approaches, while being significantly less invasive on the controlled turbine. From this point of view, tilt and yaw DIPC can be considered even less invasive forms of the helix approach. These strategies are, as expected, slightly less effective in decreasing the wake deficit, but still effectively boost wake mixing to increase the energy in the wake of a turbine.


\section*{References}
\bibliographystyle{iopart-num}
\bibliography{references}

\end{document}